\documentclass[epjCONF,columns]{svjour} 
\usepackage{graphics}
\usepackage[varg]{txfonts} 
\usepackage[latin1]{inputenc}
\session-title{2011 Hadron Collider Physics symposium (HCP-2011)}
\begin{document}
\newcommand{\pt}{\ensuremath{p_{\mathrm{T}}}}
\title{Supersymmetry and Beyond Standard Model Higgs searches at ATLAS}
\author{Olga Igonkina\thanks{\email{Olga.Igonkina@cern.nospam.ch}} on behalf of the ATLAS Collaboration}
\institute{Nikhef,  Amsterdam, The Netherlands}
\abstract{ The searches for supersymmetric and beyond Standard Model Higgs boson
  with the ATLAS detector are presented. The results are based on
  integrated luminosity of 35 pb$^{-1}$ to 1.6 fb$^{-1}$ of proton-proton
  collision data recorded at a centre-of-mass energy of 7~TeV at LHC. No
  signal is observed in any of the investigated channel and exclusion
  limits on production cross-sections are given as function of Higgs boson
  mass and of minimal supersymmetric model parameters.  } 
\maketitle
\section{Introduction}
\label{intro}
Discovering the mechanism responsible for electroweak symmetry
breaking and the origin of mass for elementary particles is one of the
primary goals of the physics program of the ATLAS experiment \cite{ATLAS}.
In the Standard Model (SM), this mechanism requires the existence of
a scalar particle, the Higgs boson \cite{Higgs}, which couples both
to bosons and fermions. In several {\it fermiophobic} extensions of SM
 \cite{fermiophobic} the Higgs field couplings to some or all fermion
generations are strongly suppressed or absent, altering the expected
Higgs boson production and decay characteristics.  For the {\it minimal
supersymmetric extension} to the Standard Model (MSSM) \cite{MSSM} two
Higgs doublets of opposite hypercharge are required, resulting in five
observable Higgs bosons. Three of these Higgs bosons (the CP even $h$, $H$, and the CP odd $A$)
are electrically neutral, while two are charged ($H^{\pm}$).
Further,  the {\it left-right symmetric}
models \cite{left-right}, the  {\it  Higgs} 
{\it triplet} models \cite{triplet} and the
{\it little Higgs} models \cite{littleHiggs} predict existence of doubly
charged Higgs bosons. In this paper, we report ATLAS results of
searches for Higgs bosons in scenarios beyond the Standard Model.

\section{Fermiophobic Higgs boson}

\begin{figure}
\resizebox{0.85\columnwidth}{5.5cm}{%
  \includegraphics{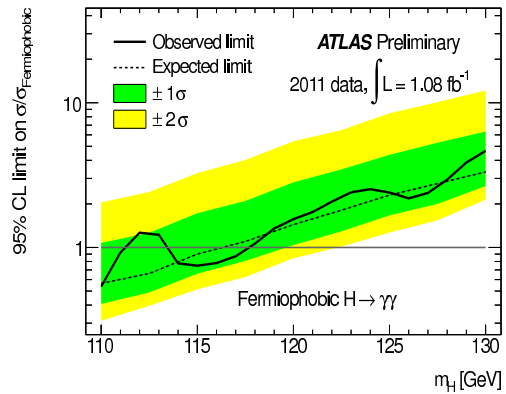} }
\caption{Exclusion limits for a fermiophobic Higgs boson normalized to the
  fermiophobic cross-section times branching ratio expectation as a
  function of the Higgs boson mass hypothesis \cite{Hgg}.  The integrated luminosity of 1.08~fb$^{-1}$ is used. }
\label{fig:Hgg}       
\end{figure}

The fermiophobic Higgs boson production and decay are characterized by
absence of coupling to the fermions. In this case, the decay of
$H\to\gamma\gamma$ is strongly enhanced, in particular for low Higgs boson
masses ($m_H$). The analysis follows the ATLAS Standard Model
$H\to\gamma\gamma$ search \cite{HggSM}, asking for two energetic photons
with transverse momenta \pt{} greater than 40 and 25 GeV, respectively.
Then the di-photon invariant mass is analysed in three ranges of the transverse
momentum of the photon pair.  The three
ranges are fitted simultaneously for Higgs boson mass hypothesis in the range
of 110~GeV to 130~GeV \cite{Hgg}. The integrated luminosity of
1.08~fb$^{-1}$ is used. The background consists mainly of 
di-photon production and misidentified photon-jet events.  No significant excess is observed as shown in the exclusion limits presented in
Fig.~\ref{fig:Hgg}. There the Higgs boson cross-section is calculated according to
\cite{fermiophobic}. The mass ranges 110-111~GeV and 113.5-117.5~GeV are
excluded at 95\% CL~\footnote{The modified frequentist method \cite{CLs}
  for upper limit calculations $CL_s$ is used for all results reported
  here.}.

\section{MSSM Higgs bosons searches}
In the MSSM, the strength of the effective Higgs boson couplings to
fermions and gauge bosons is different from those in the Standard
Model. The Higgs boson production proceeds mainly via gluon fusion or in
association with $b$ quarks, where the latter is enhanced by $\tan^2\beta$
and thus becomes more important for large values of $\tan\beta$.  While
decays to $ZZ$ or $WW$ are dominant in the Standard Model for Higgs boson
masses above 140~GeV, in the MSSM these decay modes are either suppressed
by $\cos^2(\beta-\alpha)$, where $\alpha$ is the mixing angle of the two
$CP$-even Higgs bosons, for the $H$ boson or even absent for the $A$ boson. For $h$ boson the coupling is reduced by $\sin^2(\beta-\alpha)$.
In particular, the couplings of the Higgs bosons to the third generation
down-type fermions is strongly enhanced for large region of the MSSM
parameters space.  Hence the decays of the neutral Higgs bosons into a pair
of tau leptons and decays of the charged Higgs boson to tau lepton and tau
neutrino are ones of the most promising channels. The difficult part is the
reconstruction of the tau lepton decay products, which have lower \pt{} due
to escaping neutrinos and suppression of the jet background to hadronic tau
decays ($\tau_h$). In case of leptonic tau decays, the electrons and muons
from tau typically have lower momenta than prompt leptons, and
therefore are difficult to trigger.

\subsection{Neutral MSSM $H\to\tau\tau$}
The ATLAS search for $A/H/h\to\tau\tau$ \cite{Htautau} is done in four
final states according to decays of tau leptons: $e\mu 4\nu$, $e\tau_{h} 3
\nu$, $\mu\tau_h 3 \nu$, $\tau_h\tau_h 2\nu$ using 1.06~fb$^{-1}$, where
$\tau_h$ is a hadronic tau decay.  These final states have branching ratios of
6\%($e\mu$), 23\%($\mu\tau_h$), 23\%($e\tau_h$) and 42\%($\tau_h\tau_h$).
To compensate for the neutrinos which escape undetected, various estimates
of the invariant mass are used.  For $\tau_h\tau_h$ channel the invariant mass
of the detected products, {\it visible mass}, is used. The {\it effective}
mass calculation is used in the $e\mu$ channel, where missing momentum is added
to visible decay products of taus.  In case of lepton and $\tau_h$
combination the {\it missing mass calculator} technique pioneered at
Tevatron is deployed, see \cite{MCC} for details.
The dominant background is $Z$ boson production in the channels
with an electron or a muon in the final state, while $\tau_h\tau_h$ is mainly
polluted by events with many jets (QCD) as well as events with $Z$ boson and $W$+jets.

\begin{figure}
\resizebox{0.85\columnwidth}{5.5cm}{%
  \includegraphics{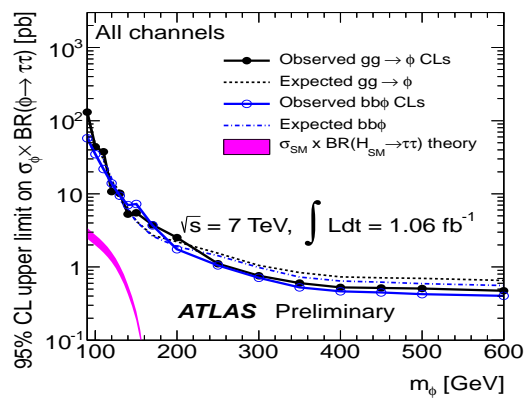} }
\caption{Expected and observed limits at 95\% CL on the production
  cross section times branching ratio for a generic Higgs boson \cite{Htautau}.}
\label{fig:Htta}       
\end{figure}

\begin{figure}
\resizebox{0.85\columnwidth}{5.5cm}{%
  \includegraphics{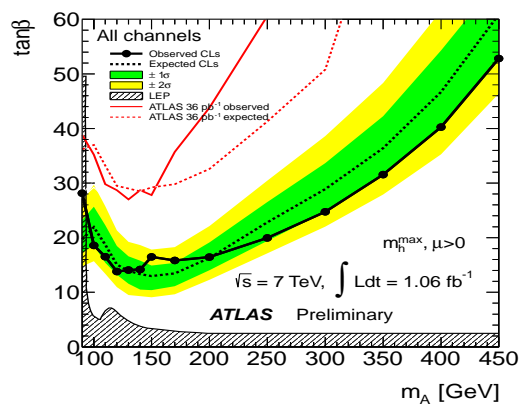} }
\caption{Expected and observed exclusion limits in the $m_A-\tan \beta$ plane of
  the MSSM derived from the combination of the analyses for the $e \mu$, $l
  \tau_h$ and $\tau_h \tau_h$ final states. The region above the drawn
  limit curve is excluded at the 95\% confidence level. \cite{Htautau}.}
\label{fig:Httb}       
\end{figure}

No signal is observed in data of 1.03~fb$^{-1}$ in any of the channels and
the exclusion limits at the 95\% CL are set on the production cross-section
times branching ratio of a generic Higgs boson $\phi$ as a function of its
mass, see Fig.~\ref{fig:Htta}. The cross-sections are calculated according to \cite{HiggsXS}.
  The cross section limit is evaluated for
signal acceptances of two different production processes, $gg\to \phi$ and
b-quark associated production which can differ from the SM Higgs boson by
the coupling strength.  The $e\tau_h$, $\mu\tau_h$ final states provide the
most stringent limit over a large part of the accessible Higgs boson mass
range. The $e\mu$ and $\tau_h\tau_h$ final states lead to improvements of
the exclusion limits for small and large Higgs boson masses, respectively.
For the MSSM the limits on the production of neutral Higgs bosos $A/H/h$ are
set as a function of the parameters $m_A$ and $\tan\beta$ for the
$m_h^{max}$ scenario and Higgsino mass parameter $\mu>0$, see
Fig.~\ref{fig:Httb}.

\subsection{Charged MSSM $H\to\tau\nu$}

The charged Higgs boson with mass lower than the mass of the top quark is
searched for in top quark pair production where one of top quark decaying as
$t\to H^+ b, H\to \tau\nu$. The following final states are considered :
\begin{enumerate}
\item Both tau and second top quark decaying hadronically \cite{Htnh},
\item Both tau and second top quark decaying leptonically \cite{Htnl},
\item tau decaying leptonically and second top quark decaying hadronically \cite{Htnl}.
\end{enumerate}

In first case, the signal is searched for using the transverse mass of the tau
candidate and the missing momentum \cite{Htnh}.  The dominant top quark and
$W$+jet backgrounds are estimated using event topology similar to the signal,
but with muon instead of the tau in the final state. The reconstructed muon
is then replaced by simulated tau decaying hadronically, where tau momentum
set equal to the reconstructed muon momentum. The multi-jet background is
estimated using control region by reverting tau identification and b-tagging requirements.

In two latter cases, the signal is searched for using the invariant mass of $b$-quark and electron or muon
coming from the same top-quark and a Higgs boson candidate transverse mass
by performing maximization of the invariant mass over possible values for
neutrino momenta in each event \cite{Htnl}. These variables allow
additional discrimination over the backgrounds consisting mainly of top
quark pairs. In both channels the data agree well with Standard Model
expectations.

\begin{figure}
\resizebox{0.85\columnwidth}{5.5cm}{%
  \includegraphics{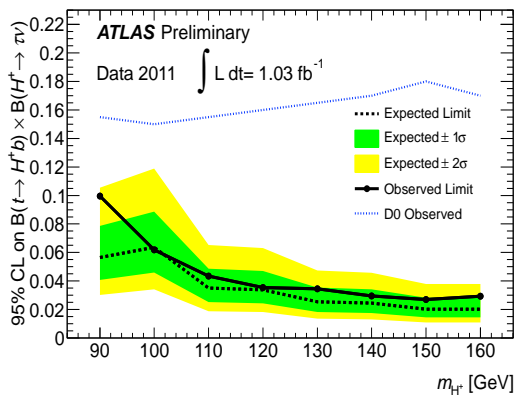} }
\caption{ Expected and observed 95\% CL exclusion limits for charged Higgs
  boson production from top quark decays hadronic tau decays in final
  state, obtained for integrated luminosity of 1.03~fb$^{-1}$.  Results are
  shown as a function of $m_{H^+}$ in terms of BR$(t\rightarrow H^+b)$
  $\times$ BR$(H^+\rightarrow\tau^+\nu)$ \cite{Htnh}.}
\label{fig:Htnh1}       
\end{figure}

\begin{figure}
\resizebox{0.85\columnwidth}{5.5cm}{%
  \includegraphics{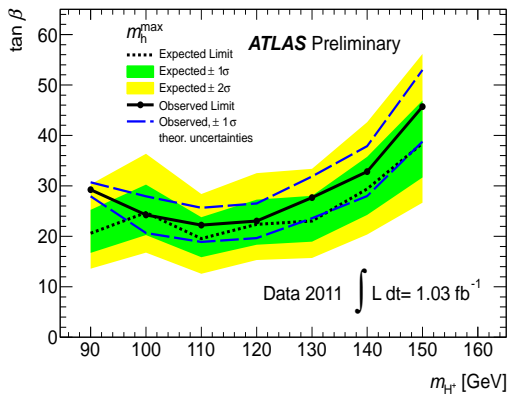} }
\caption{Upper limits for charged Higgs boson production from top quark
  decays with hadronic tau decays in final state, obtained for integrated
  luminosity of 1.03~fb$^{-1}$.  Results are shown for the MSSM scenario
  $m_h^{\mathrm{max}}$ in the $m_{H^+}$-$\tan \beta$ plane \cite{Htnh}.}
\label{fig:Htnh2}       
\end{figure}

\begin{figure}
\resizebox{0.85\columnwidth}{5.5cm}{%
  \includegraphics{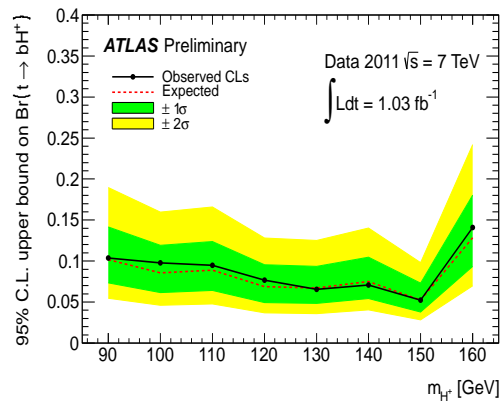} }
\caption{ Upper limits on $Br(t \to bH^+)$ for the combined single-lepton
  and di-lepton channels, as a function of the charged Higgs boson mass,
  obtained for an integrated luminosity of 1.03/fb and with the assumption
  that $Br(H+ \to \tau\nu) = 1$ \cite{Htnl}. }
\label{fig:Htnl1}       
\end{figure}

\begin{figure}
\resizebox{0.85\columnwidth}{5.5cm}{%
  \includegraphics{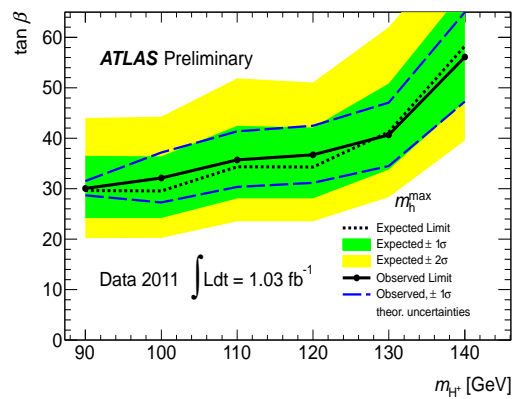} }
\caption{Upper limits for charged Higgs boson production from top-quark
  decays in the $m_{H^+}$-$\tan\beta$ plane, in the context of the $m_h^{max}$
  scenario of the MSSM, obtained for an integrated luminosity of 1.03/fb
  and with the assumption that $Br(H^+ \to \tau^+\nu) = 1$ \cite{Htnl}.}
\label{fig:Htnl2}       
\end{figure}

No excess is observed in 1.03~fb$^{-1}$ in either of three channels and the
upper limits on cross-section multiplied with branching ratio $Br(t\to
Hb)\cdot Br(H\to\tau\nu)$ are set as shown in Fig.~\ref{fig:Htnh1} and
\ref{fig:Htnl1}. The upper limits for charged Higgs boson production from top decays
as function of MSSM parameters $m_A$ and $\tan\beta$ are given on
Fig.~\ref{fig:Htnh2} and \ref{fig:Htnl2}.  Although the presence of leptons
in final state allows to suppress backgrounds significantly, the statistics
in case of hadronic tau decay is higher, which results in lower upper
limits. The combination of all three channels is not yet available.



\subsection{Charged MSSM $H^+\to c\bar{s}$}

\begin{figure}
\resizebox{0.85\columnwidth}{5.5cm}{%
  \includegraphics{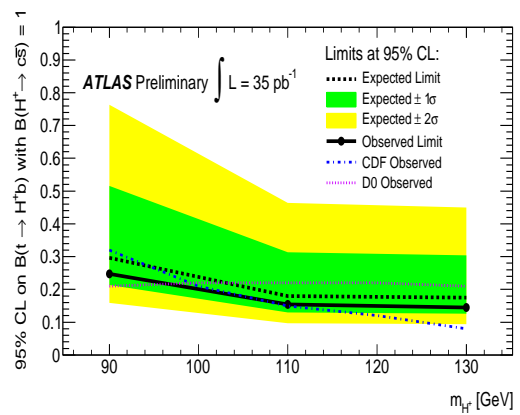} }
\caption{ The extracted 95\% C.L. upper limits on $B(t \to H+b)$ from the
  ATLAS data of integrated luminosity of 35 pb$^{-1}$ \cite{Hcs} are compared with the
  expected results and results from the Tevatron. The results assume $Br(H+
  \to c\bar{s})=1$.}
\label{fig:Hcs1}       
\end{figure}

The ATLAS Collaboration has also performed the search of charged Higgs
boson in $H^+\to c\bar{s}$ channel \cite{Hcs}. Such decay is an important mode at low
values of $\tan\beta < 1$ with $Br(H^+\to c\bar{s})$ near 40\% for $m_{H^+}
\simeq 130$~GeV. The electron or muon decays of the second top quark are
selected. The signal has the same characteristics as semi-leptonic
$t\bar{t}$, with the exception of the mass of the two jets from $H^+$,
which will peak at $m_{H^+}$, rather than $m_W$. Both the overall number of
events observed and the shape of the dijet mass spectrum is analysed. The
kinematic fit of the events is performed to select jets originating from the
Higgs boson.  The number of observed events and dijet mass distribution is found
in good agreement with the expectations from the SM and the limits set on
the branching ratio $Br(t \to H^+b)$, assuming $Br(H^+ \to c \bar{s}) =
1$, see Fig.~\ref{fig:Hcs1}. This result can be used to set limits for an anomalous scalar charged
boson decaying to dijets in top quark decays, as no model-specific
parameters are used in this analysis.

\section{Doubly  Charged Higgs boson}

\begin{figure}
\resizebox{0.85\columnwidth}{5.5cm}{%
  \includegraphics{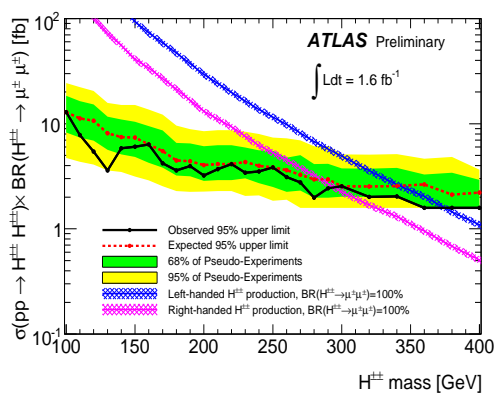} }
\caption{ Upper limit at 95\% CL on the cross section times branching ratio
  for pair production of doubly charged Higgs bosons decaying to two muons
  using 1.6~fb$^{-1}$ of data \cite{Hpp}. Superimposed is the predicted cross section
  for left-handed and right-handed doubly charged Higgs boson production assuming
  the branching ratio to muons is 100\%. }
\label{fig:Hpp}       
\end{figure}

Doubly charged Higgs bosons are predicted by a number of models such as
left-right symmetric models \cite{left-right}, Higgs triplet \cite{triplet}
or little Higgs models \cite{littleHiggs}. The dominant production process
of doubly charged Higgs boson is a pair production via the Drell-Yann
process $pp\to H^{++}H^{--}$. The search is performed in events with two
muons with the same charge using 1.6~fb$^{-1}$ \cite{Hpp}.  Only coupling
values between $10^{-5}$ and 0.5 of the $H^{\pm\pm}$ to muons are
considered to ensure a short lifetime ($c\tau<10 \mu\rm{m}$) and to ensure
that the relative natural width of the Higgs boson is less than 1\%.  The
data are in good agreement with the background prediction. The backgrounds
at low masses are populated by non-prompt muons, originating in heavy
flavor decay or decay-in-flight of pions or kaons. The high mass range has an
additional contribution from di-boson production.  No evidence for any
resonant production of like-sign di-muon pairs is observed. Stringent limits are placed on $H^{\pm\pm}$ masses
between 295 GeV (375 GeV) for the right-handed (left-handed) production
assuming $Br(H^{\pm\pm}\to\mu^{\pm}\mu^{\pm}) = 1$, see
Fig.~\ref{fig:Hpp}. The details on cross-section calculations could be
found in \cite{Hpp}. Assuming branching ratio of 33\% the corresponding
limits become 210 (268)~GeV, respectively.

\end{document}